\documentclass[aps,prl,twocolumn,showpacs]{revtex4}
\usepackage{graphicx}
\usepackage{amsmath}

\include{NewCommands}

\begin{document}

\title{Nanoscale spin-polarization in dilute magnetic semiconductor (In,Mn)Sb}

\author{A.~Geresdi, A. Halbritter, M. Csontos, Sz. Csonka, G. Mih\'aly}

\affiliation{{Department of Physics, Budapest University of
Technology and Economics and \\
Condensed Matter Research Group of the Hungarian Academy of Sciences, 1111 Budapest, Budafoki ut 8., Hungary}}

\author{T. Wojtowicz}
\affiliation{Institute of Physics, Polish Academy of Sciences, PL-02668 Warsaw, Poland}

\author{X. Liu}
\author{B. Jank\'o}
\author{J.K. Furdyna}
\affiliation{Department of Physics, University of Notre Dame, Notre Dame, Indiana 46556, USA}

\date{\today}

\begin{abstract}
Results of point contact Andreev reflection (PCAR) experiments on (In,Mn)Sb are presented and analyzed in terms
of current models of charge conversion at a superconductor-ferromagnet interface. We investigate the influence
of surface transparency, and study the crossover from ballistic to diffusive transport regime as contact size is
varied. Application of a Nb tip to a (In,Mn)Sb sample with Curie temperature $T_\textrm{C}$ of $5.4\,$K allowed
the determination of spin-polarization when the ferromagnetic phase transition temperature is crossed. We find a
striking difference between the temperature dependence of the local spin polarization and of the macroscopic
magnetization, and demonstrate that nanoscale clusters with magnetization close to the saturated value are
present even well above the magnetic phase transition temperature.
\end{abstract}

\maketitle

Controlling the spin state of electrons provides an important versatility for future electronics
\cite{Awschalom}. Most of the envisioned spintronic devices are based on spin transfer mechanisms on the nanoscale.
For this purpose new materials have been synthesized with highly spin polarized bands, and novel experimental
techniques are being applied to characterize the spin state of the charge carriers \cite{Soulen_Ji_Upadhyay,
Strijkers}.

(III,Mn)V dilute magnetic semiconductors are promising spintronic materials with high spin polarization
\cite{Braden,Panguluri,Panguluri_2007} and with a wide variety of spin-dependent transport properties
\cite{Ohno}. While considerable effort is concentrated to enhance the ferromagnetic transition temperature
\cite{Jungwirth,Jamet}, studies of low $T_\textrm{C}$ alloys are also of great interest, as they contribute to a
better understanding of the underlaying physics. Here the alloy (In,Mn)Sb -- with Curie temperatures below the
transition temperatures of conventional superconductors -- is especially interesting in that it allows one to
study the spin polarization by Andreev spectroscopy \textit{both} in the ferromagnetic and paramagnetic phases.

The Andreev reflection experiment provides a direct measure of the \textit{current spin polarization}, $P$. The
current through a ferromagnet/superconductor interface is determined by the charge conversion of individual
electrons to Cooper pairs. As a Cooper pair consists of two electrons with opposite spins, the conversion is
suppressed in case of spin polarized bands, so that $P$ can be deduced from the voltage dependence of the
conductance. $P$ is often derived in the framework of the modified BTK theory \cite{Strijkers}, which simply
splits the current to unpolarized and fully polarized parts. The net current is then calculated as \(
I_\textrm{total}=(1-P_\textrm{BTK})I_\textrm{unpol}+P_\textrm{BTK}I_\textrm{pol} \) by assuming no Andreev
reflection for the fully polarized current and applying the original BTK theory for the unpolarized part
\cite{BTK}. An alternative, more rigorously founded quantification of $P$ can be obtained based on the imbalance
of spin-dependent transmission coefficients
$P_\textrm{T}=(T_{\uparrow}-T_{\downarrow})/(T_{\uparrow}+T_{\downarrow}$) \cite{Cuevas,Lohneysen}. Both models
assume ballistic transport, but due to the difference in the approaches they cannot be mapped to each other
mathematically.

In this paper point contact Andreev reflection (PCAR) spectra are presented for (In,Mn)Sb films with
ferromagnetic transition temperature of $T_\textrm{C}=5.4\,$K, and for its (In,Be)Sb non-magnetic counterpart.
We analyze the data in terms of the above models for various surface barriers, and show that the deduced
spin-polarizations agree well for the transparent contact limit. We also investigate the crossover from the
ballistic to the diffusive transport regime as the contact diameter is varied in a controlled manner.
Furthermore the PCAR experiments allow us to compare the temperature dependence of the measured local spin
polarization to that of the macroscopic magnetization as the ferromagnetic phase transition temperature is
crossed.

Thin In\(_\textrm{1-x}\)Mn\(_\textrm{x}\)Sb  and In\(_\textrm{1-y}\)Be\(_\textrm{y}\)Sb film with typical
thicknesses of $200\,$nm were grown by molecular beam epitaxy in a Riber 32 R\&D system, and were characterized
by structural, transport and magnetic measurements \cite{Wojtowicz}. The hole concentration of these samples is
$n\approx 2 \cdot 10^{20}\,$cm$^{-3}$, resulting in a metallic conductivity with $\sigma \approx 3 \cdot 10^3\,\Omega ^{-1}$cm$^{-1}$. In the PCAR experiments mechanically-edged Nb tips were used as the superconducting
electrodes. The position of the tip was regulated by a screw mechanism and a piezo actuator. The accuracy of the
positioning is $0.1\,$nm, as determined from currents measured in the tunneling regime, i.e. before touching the
sample surface. In the present study the voltage-dependent differential conductance was acquired using standard
four-probe measurements, applying noise filters in the low temperature stage of the sample holder.

\begin{figure}[!t]
\includegraphics[width=0.8\columnwidth]{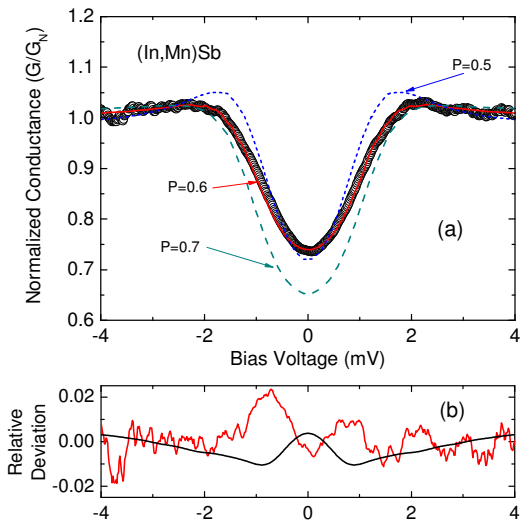}
\caption{\textit{(Color online) a) Normalized conductance of a Nb-(In,Mn)Sb contact at $T=4.2\,$K, with fits
using the BTK method. The red curve with \(P=0.60 \pm 0.01\) yields the best fit (the other fitting parameters
are: $Z=0.13$, $T=4.17\,$K, $\Delta=1.13\,$meV). The dashed lines are fits with intentionally detuned
polarizations (P=0.5, P=0.7) using the same temperature and gap parameters as for the best fit, and with Z as a
fitting parameter.  b) Red curve: deviation of measured data and BTK fitting; black curve: deviation between the
two fitting methods as discussed in the text.}} \label{FigFixedPFits}
\end{figure}

A typical PCAR spectrum is shown in Fig.~\ref{FigFixedPFits} for (In,Mn)Sb at $T=4.2\,$K. The bias dependence of
the normalized conductance was analyzed in terms of the two models discussed earlier. The best fit obtained with
the modified BTK model is shown by the red curve in Fig.~\ref{FigFixedPFits}(a). This represents an almost
transparent contact (the $Z=0.13$ barrier strengths corresponds to $T=1/(1+Z^2)=0.98$ transmission) and high
spin polarization \(P_\textrm{BTK}=0.60\pm 0.01 \), in good agreement with earlier experimental data
\cite{Panguluri}. It is to be noted that simulated curves for spin polarization of \(0.50\) or \(0.70\) are far
away from the measured data, i.e. the value of \(P_\textrm{BTK}\) can really be determined with a high accuracy
within this formalism.

Similar high quality fit can also be obtained by calculating the imbalance of spin-dependent transmission
coefficients, $P_\textrm{T}$ \cite{Cuevas,Lohneysen}. There is a small but clear systematic deviation between
the two fitting methods, as expected due to differences in the two formalisms [see Fig.~\ref{FigFixedPFits}(b)].
However, the difference between simulations based on the two methods is within the scatter of experimental data
and -- surprisingly -- the fitting parameters for the transmissions \(T_\uparrow=0.99\) and
\(T_\downarrow=0.246\) obtained by this approach correspond to a polarization of \(P_\textrm{T}=0.605\), which
agrees very well with that derived by the BTK theory. A more detailed analysis was carried out by acquiring
data using several different contacts. We conclude that, despite the fact that the two models are based on quite
different assumptions and cannot be mapped onto each other, they lead to identical results for high quality
transparent contacts (denoted by \(Z \rightarrow 0\) and \(T_\uparrow \rightarrow 1\)). For less transparent
contacts, i.e. for $Z > 0.3$, which correspond to $T < 0.9$, the fitting curves obtained using the two methods
are still almost identical, but the above excellent agreement in the deduced polarization is lost. Below we
present the BTK analysis of a large set of data obtained on various samples by contact formation at different
positions on the sample surface.

\begin{figure}[!b]
\includegraphics[width=0.8\columnwidth]{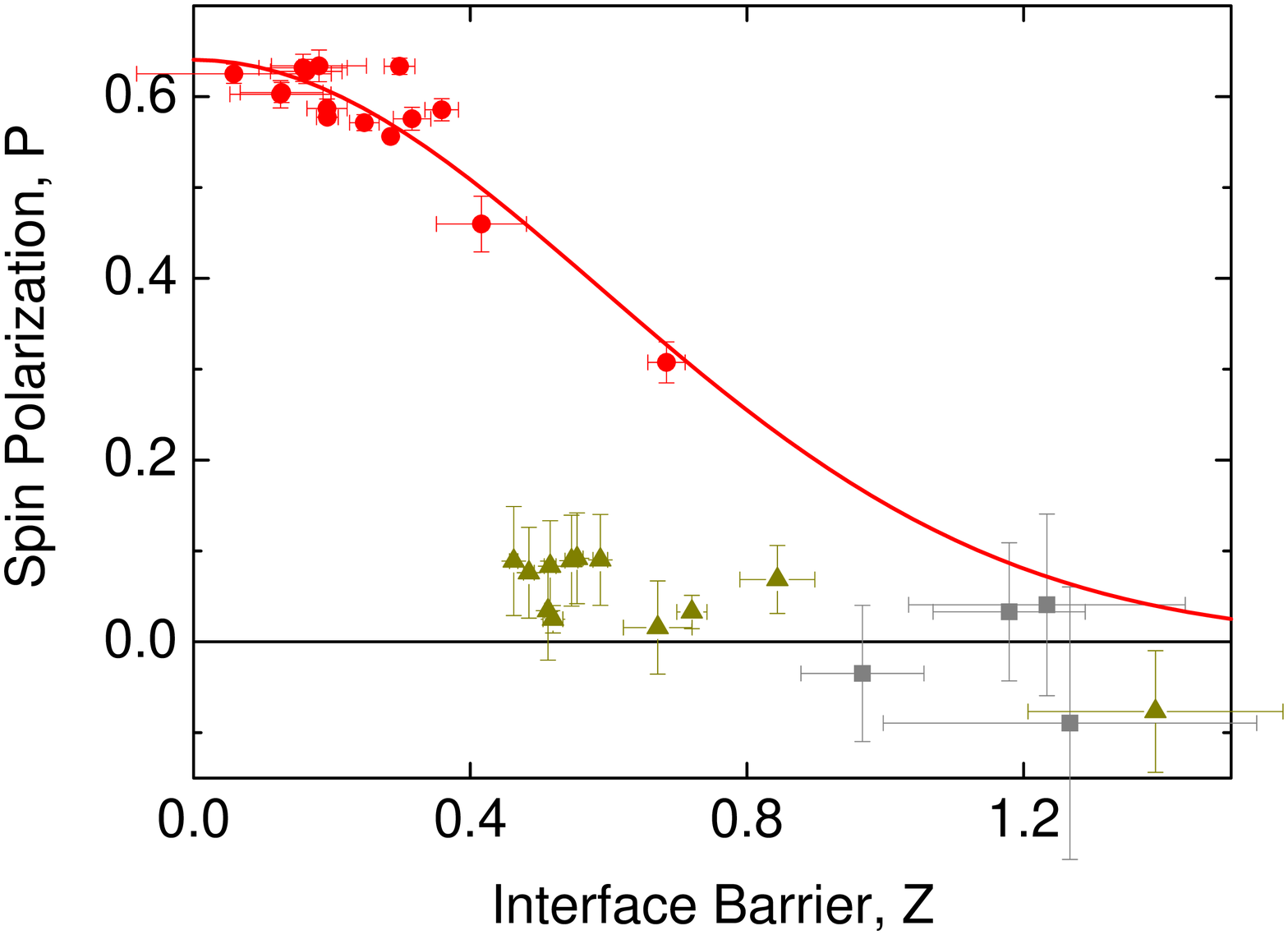}
\caption{\textit{(Color online) Experimental results for several contacts on the P-Z plane. Extracted
polarization data for Nb-(In,Mn)Sb contacts are denoted by red circles, while the solid red line represents a
Gaussian fit based on Ref.~\cite{Kant}. Reference data on Nb-(In,Be)Sb and Nb-Au contacts are shown by gray
squares and green triangles, respectively.}} \label{FigPZPlane}
\end{figure}

We display our results using several contacts with various transparencies in the standard way.
Figure~\ref{FigPZPlane} shows the fitting parameters on the P-Z plane both for the magnetic semiconductor
(In,Mn)Sb and for its non-magnetic counterpart, (In,Be)Sb. Test results on a simple paramagnetic metal (gold) are
also shown. The decay of spin-polarization with increasing barrier strength observed with Nb-(In,Mn)Sb
contacts is attributed to spin-flip scattering in the contact area, and the intrinsic spin polarization of the
sample is deduced from fitting the data to a Gaussian shape \cite{Kant}, $P(Z \rightarrow 0) =0.62\pm 0.01$. In contrast, the Nb-Au and the Nb-(In,Be)Sb contacts do not exhibit a finite spin polarization, as expected. It is
worth noting that the accuracy of the polarization determined from an individual measurement is reduced
for high-Z contacts.

\begin{figure}[!t]
\includegraphics[width=0.95\columnwidth]{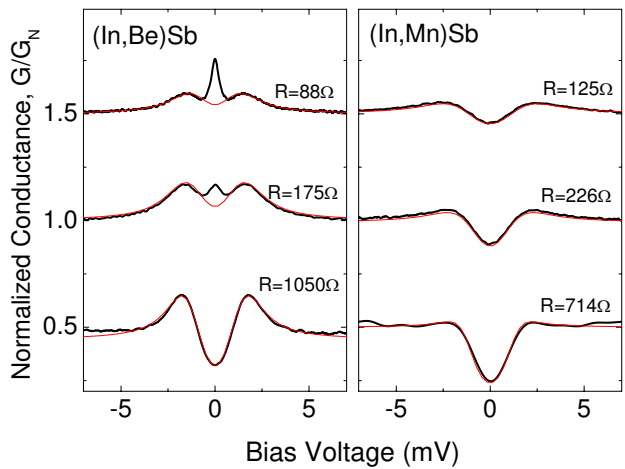}
\caption{\textit{Normalized conductance of several contacts with different contact resistances for nonmagnetic
(In,Be)Sb samples (left panel) and magnetic (In,Mn)Sb samples (right panel). The experiments were performed at
$T=4.2\,$K. The fits based on modified BTK-theory assuming finite lifetime ($\Gamma>0$) are shown in the red,
the curves are shifted vertically for clarity.}} \label{FigDiffusivePeak}
\end{figure}

The high-accuracy piezo positioning of the Nb tip allows controlled variation of the contact diameter in the
range of about 5~to~50\,nm. The Andreev spectra of the magnetic and nonmagnetic samples prepared by the same MBE
technique and having nearly identical bulk parameters \cite{Wojtowicz} are shown in Fig.~\ref{FigDiffusivePeak}.
In these experiments the contact diameter was increased above the literature value of the heavy hole mean free
path $l_\textrm{m} \approx 15\,$nm, i.e. to the region where diffusive transport is expected. The contact size
was estimated from the quasiclassical formula of contact resistance, applicable both to ballistic and diffusive
transport \cite{Nikolic}:
\begin{equation}
R=\left (1+Z^2\right)\left(\frac{4\rho l_\textrm{m}}{3 \pi d^2}+\gamma\left(\frac{l_\textrm{m}}{d} \right)
\frac{\rho}{2d}\right),
\end{equation}
where \(d\) is the diameter of the contact, \(\rho\) is the bulk resistivity of the material
\cite{Wojtowicz,Vurgaftman}, and \(\gamma\) is a prefactor of the order of unity.

For large contact areas (small resistances), an unambiguous qualitative feature of diffusive transport is the
narrow zero-bias peak observed in the nonmagnetic (In,Be)Sb sample (Fig.~\ref{FigDiffusivePeak}, left panel).
This is attributed to multiple phase-coherent reflections occurring when $d > l_\textrm{m}$ and the phase-coherence length $l_\phi > d$, similarly to the reflectionless tunneling phenomenon observed in superconductor normal metal tunnel junctions \cite{Kastalsky, Beenakker}. In our case the peak is broadened by the thermal energy corresponding to about $350\,\mu$V. A detailed quantitative description of multiple reflections in diffusive contact regime based on scattering matrix calculations is given elsewhere \cite{Geresdi}.

In general, such zero-bias coherence peak is not expected if the Cooper pair conversion occurs in a magnetic
sample where the energy of the resulting quasiparticles of opposite spins differs due to the exchange splitting.
In that case the phase coherence is lost within a characteristic time determined by this energy difference,
$t_\phi=\hbar /\Delta E_\textrm{m}$. Indeed, no peak is observable for the Nb-(In,Mn)Sb contact, as shown in the
right panel of Fig.~\ref{FigDiffusivePeak}. Note, however, that a simple mean field approach for the local
magnetic interaction, $\Delta E_\textrm{m} \approx k_\textrm{B}T_\textrm{C}$, would mean only a slight
broadening of the peak instead of its complete suppression, since in our case $k_\textrm{B}T_\textrm{C}$ is
almost as small as the thermal energy at liquid helium temperature at which the experiment was performed
($T_\textrm{C}=5.4\,$K). Consequently, the absence of the coherence peak implies a much more radical reduction
of phase coherence time, i.e., that the magnetic splitting tested on the length scale of a few nm is $\Delta
E_\textrm{m} \gg k_\textrm{B}T_\textrm{C}$.

\begin{figure}[!h]
\includegraphics[width=0.8\columnwidth]{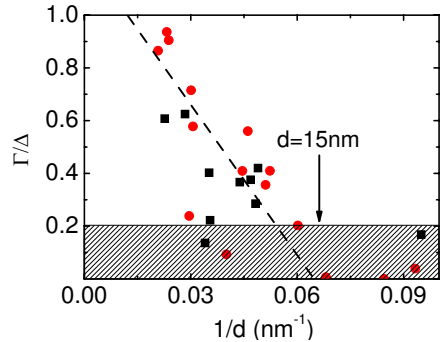}
\caption{\textit{(Color online) Normalized quasiparticle lifetime parameter as a function of contact size. Red
circles and black squares denote data acquired on InMnSb and InBeSb samples respectively. The onset of finite
damping appears at \(d \approx 15\,\textrm{nm}\). The dashed line is a guide for the eye.}}
\label{FigDiffusiveOnset}
\end{figure}

Another feature of low-resistance contacts is the smearing of the Andreev spectra on larger voltage scales, both
for the magnetic and for nonmagnetic samples. We have found that for contacts with $d\gtrsim15$~nm the BTK
theory gives unphysical parameters. The fitting temperature obtained by this approach is above the
superconducting transition temperature of Nb, while the value of the superconducting gap $\Delta$ still
corresponds to that measured at liquid helium temperature. If, on the other hand, the temperature is fixed at
the correct value, the BTK theory gives a rather poor fit. This broadening, however, can be taken into account
by introducing a finite quasiparticle lifetime (denoted by \(\Gamma\)) on the superconductor side
\cite{Plecenik}. The physical meaning of this phenomenological parameter is the enhanced probability of
inelastic scattering in the diffusive regime. Plotting the dimensionless quasiparticle lifetime parameter
\(\Gamma/\Delta\) as a function of the contact diameter, we also see that the onset of the diffusive process
appears around \(d \approx 15\,\)nm, as shown in Fig.~\ref{FigDiffusiveOnset}. This feature is independent on
whether the sample is magnetic or not: typical fits to data are shown in Fig.~\ref{FigDiffusivePeak}. We
conclude that the smoothing can be attributed to diffusive scattering in the contact area, and that it
disappears when the contact size is below the mean free path.

\begin{figure}
\includegraphics[width=0.8\columnwidth]{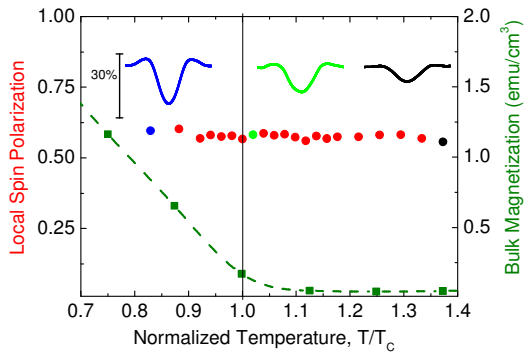}
\caption{\textit{(Color online) Temperature dependence of the spin-polarization acquired from point contact
measurement (circles); and the remanent magnetization determined by SQUID measurement (squares). The PCAR
results are reproducible for contacts prepared at various surface positions.}} \label{FigPvsTemperature}
\end{figure}

We have also investigated the temperature dependence of the Andreev spectra of (In,Mn)Sb. This study is
especially interesting because the use of Nb tip allowed us to cross the ferromagnetic phase transition
temperature. The results are summarized in Fig.~\ref{FigPvsTemperature}. Although temperatures much above the
Curie-temperature had been reached, only thermal broadening is observed, and the dip in the differential
conductance remains clearly present. In the analysis of the curves the temperature was used as a fitting
parameter, and the values deduced from the measured Andreev spectra agree within \(0.1\,\)K with the actual
temperature measured independently. One of the most important result of the present study is that \textit{the
spin polarization extracted from the fits does not vanish in the paramagnetic phase}. For comparison the
remanent magnetization measured by SQUID on the same sample is also shown in Fig.~\ref{FigPvsTemperature}.

The above surprising behavior directly confirms the percolation nature of the magnetic phase transition in
dilute magnetic semiconductors \cite{MacDonald}. The macroscopic magnetization signifies the ordering of
magnetic clusters at the Curie temperature, while the local measurement of spin polarization on the $10\,$nm
length scale reveals finite magnetization at temperatures as high as 40 \% above the phase transition. Moreover,
the magnetization of the individual clusters does not show any significant change at $T_\textrm{C}$, as
reflected in the temperature-independent spin polarization. This implies that the characteristic energy scale of
cluster formation is much higher than $k_\textrm{B} T_\textrm{C}$, in agreement with our earlier analysis of the
absence of coherence effects in (In,Mn)Sb. The fact, that magnetic clusters with nearly saturated magnetization
are present well above $T_\textrm{C}$ may open the possibility of nanoscale spintronic applications at
temperatures far above those required for the \textit{macroscopic} magnetic ordering.

In conclusion, point contact Andreev-reflection experiments were performed on (In,Mn)Sb and (In,Be)Sb under
various circumstances. Results obtained on samples with different surface barriers were analyzed in terms of the
extended BTK theory and of the ``spin-dependent transmission model''; and it was shown that for the transparent
contact limit the two formalisms lead to identical spin polarization: $P=0.61 \pm 0.01$. By increasing the
contact size in a controlled manner, we were able to enter from the ballistic to the diffusive transport regime,
where zero bias peak due to multiple phase-coherent reflections and smearing of the Andreev spectra were observed. Analyzing the quasiparticle lifetime in (In,Mn)Sb, we found that reliable experiments in the ballistic limit can only be obtained if the contact diameter is less than \(15\,\)nm, that corresponds to the heavy hole mean free path. The
temperature dependence of the spin-polarization P was also investigated, and a striking difference was found
between P and the remanent magnetization. Our observation directly confirms the percolation scheme of the phase
transition, with clusters characterized by nearly saturated magnetization even well above the magnetic phase
transition temperature.

This research was supported by the National Science Foundation Grants DMR 02-10519 and
DMR 06-3752; NSF-NIRT award ECS-0609249; US. Department of Energy, Basic Energy
Sciences contract W-31-109-ENG-38; and by the Hungarian Scientific Research Fund OTKA under Grant NK72916 and F49330. A. Halbritter is a grantee of the Bolyai J\'anos Scholarship.

\end{document}